\documentclass[preprint,showpacs,preprintnumbers,amsmath,amssymb]{revtex4}

\usepackage{graphicx}
\usepackage{dcolumn}
\usepackage{bm}
\usepackage{epsfig}

\begin{document}
\title{Yang-Yang method for the thermodynamics of one-dimensional multi-component interacting fermions}

\author{J.Y. Lee$^{1}$,
 X.W. Guan$^{1,\ddagger}$
 and M.T. Batchelor$^{1, 2}$}

 \affiliation{$^{1}$ Department of Theoretical Physics,
 Research School of Physics and Engineering,
 Australian National University, Canberra ACT 0200, Australia}

 \affiliation{$^{2}$ Mathematical Sciences Institute,
 Australian National University, Canberra ACT 0200, Australia}

\date{\today}

\begin{abstract}
Using Yang and Yang's particle-hole description, we present a
thorough derivation of the thermodynamic Bethe ansatz equations for
a general $SU(\kappa)$ fermionic system in one-dimension for both
the repulsive and attractive regimes under the presence of an
external magnetic field. These equations are derived from
Sutherland's Bethe ansatz equations by using the spin-string
hypothesis. The Bethe ansatz root patterns for the attractive case
are discussed in detail. The relationship between the various phases
of the magnetic phase diagrams and the external magnetic fields is
given for the attractive case. We also give a quantitative
description of the ground state energies for both strongly repulsive
and strongly attractive regimes.
\end{abstract}



\maketitle

\section{Introduction}

Exactly solvable models of interacting fermions in one-dimension
(1D) have attracted theoretical interest for more than half a
century. Before 1950, it was not clear how to treat the
Schr\"{o}dinger equation for a large system of interacting fermions.
The first important breakthrough was achieved by Tomonaga
\cite{Tomonaga1950} who showed that fermionic interactions in 1D can
mediate new collective degrees of freedom that are approximately
bosonic in nature.

In 1963, Luttinger \cite{Luttinger1963} introduced an exactly
solvable many-fermion model in 1D which consists of two types of
particles, one with positive momentum and the other with negative
momentum. However, Luttinger's model suffers from several flaws
which include the assumption that the fermions are spinless and
massless, and more importantly an improperly filled negative energy
Dirac sea. Mattis and Lieb \cite{Mattis1965} expanded on Luttinger's
work by correctly filling the negative energy states with ``holes''.
Before that, Lieb and Liniger \cite{Lieb1963a,Lieb1963b} solved the
1D interacting Bose gas with $\delta$-function interactions using
Bethe's hypothesis \cite{Bethe1931}. Later McGuire
solved the equivalent spin-1/2 fermion problem for the special case
where all fermions have the same spin except one having the opposite
spin in the repulsive \cite{McGuire1965} and attractive
\cite{McGuire1966} regimes. He showed that in the presence of an
attractive potential a bound state is formed. 
Further progress by Lieb and Flicker \cite{LF} followed on the two down 
spin problem.
In 1967, Yang
\cite{Yang1967} solved the fermion problem for the most general case
where the number of spin ups and spin downs are arbitrary by making
use of Bethe's hypothesis. At the same time, Gaudin
\cite{Gaudin1967} solved this problem for the ground state with no
polarization.

Sutherland \cite{Sutherland1968} then showed that the fermion model
with a general $SU(\kappa)$ spin symmetry is integrable and the
solution is given in terms of $\kappa$ nested Bethe ansatz (BA)
equations. And in 1970, Takahashi \cite{Takahashi1970} examined the
structure of the bound states in the attractive regime with
arbitrary spin and derived the ground state energy together with the
distribution functions of bound states in terms of a set of coupled
integral equations. Using Yang and Yang's method \cite{YangYang1969}
for the boson case, Takahashi \cite{Takahashi1971} and Lai
\cite{Lai1971,Lai1973} derived the so-called thermodynamic Bethe
ansatz (TBA) equations for spin-1/2 fermions in both the repulsive
and attractive regimes. The spin-string hypothesis describing the
excited states of spin rapidities was also introduced by both
authors. Later on, Schlottmann
\cite{Schlottmann1993,Schlottmann1994} derived the TBA equations for
$SU(\kappa)$ fermions with repulsive and attractive interactions. 
See also Schlottmann's epic review article on exact results for 
highly correlated electron systems in 1D \cite{S-review}.

The TBA equations have been analyzed in several limiting cases, i.e.,
$T\rightarrow 0$, $T\rightarrow\infty$, $c\rightarrow 0$ and
$|c|\rightarrow\infty$, where $T$ is the temperature and $c$ is the
interaction strength. The ground state properties and the elemental
charge and spin excitations were also studied for some special
cases. 
However, the TBA equations for the attractive regime \cite{Schlottmann1994,S-review} 
are not the most convenient for the analysis of phase
transitions and thermodynamics. For the attractive case, it was
shown that the ground state in the absence of symmetry breaking
fields consists of spin neutral charge bound states of $\kappa$
particles. The repulsive case however consists of freely propagating
charge states and spin waves with different velocities. The
phenomenon of spin-charge separation plays a ubiquitous role in the
low energy physics of 1D systems \cite{Giamarchi}. 
However, the physics of these models, such as the universal thermodynamics of 
Tomonaga-Luttinger liquids, quantum criticality and the universal nature of contact interaction, 
are largely still hidden in the complexity of the TBA equations.
It is thus important to develop new methods to extract the physics of 1D exactly solved 
many-body systems in order to bring them more closer to experiments. 

Most recently, experimental advances in trapping and cooling atoms
to very low temperatures allow a test of the theoretical predictions
made so far. In particular, Liao {\em et al.} \cite{Liao2009}
experimentally studied  spin-1/2 fermions of ultracold $^{6}$Li
atoms in a 2D array of 1D tubes with spin imbalance. The phase
diagram was confirmed and it was discovered that a large fraction of
a  Fulde-Ferrell-Larkin-Ovchinnikov (FFLO)-like phase lies in the trapping center accompanied by two
wings of a fully paired phase or unpaired phase depending on the
polarization. This observation verified the theoretical predictions
\cite{Orso,Hu,Guan2007,Mueller} regarding the phase diagram and
pairing signature for the ground state of strongly attractive
spin-1/2 fermions in 1D.
Although the FFLO phase has not yet been 
 observed  directly, the experimental results pave the way to direct observation 
and characterization of FFLO pairing  \cite{Liao2009}.

In this paper, we derive the TBA equations for a general 1D system
of fermions with $SU(\kappa)$ spin symmetry from Sutherland's BA
equations using the same approach as Yang and Yang for 1D bosons
\cite{YangYang1969}. Both the repulsive and attractive cases are
discussed. We also give the exact thermodynamics of the ground state
of the attractive and repulsive cases in both the strong coupling
and weak coupling limits. A general relationship between the
different magnetic phases and the external magnetic field is
discussed for the attractive case. How the external magnetic fields
affect the different pairing phases in the attractive regime is also
addressed. This paper gives a thorough derivation of many results in
a recently published paper \cite{Guan2010} that provides the exact
low temperature thermodynamics for strongly attractive $SU(\kappa)$
fermions with Zeeman splitting and shows that the system behaves
like a universal Tomonaga-Luttinger liquid  in the gapless
phase.

\section{The Model}
The Hamiltonian for the 1D $N$-body problem is
\begin{equation}
H=-\frac{\hbar^{2}}{2m}\sum_{i=1}^{N}\frac{\partial^{2}}{\partial
x_{i}^{2}}+g_{1D}\sum_{1\leq i<j\leq
N}\delta(x_{i}-x_{j})+\sum_{i=1}^{\kappa}N^{i}\epsilon^{i}_{Z}(\mu_{B}^{i},B).
\end{equation}
It describes $N$ fermions of the same mass $m$ confined to a 1D
system of length $L$ interacting via a $\delta$-function potential.
The first and second terms in the Hamiltonian correspond to the
kinetic energy and $\delta$-interaction potential respectively. The
coupling constant $g_{1D}$ can be expressed in terms of the
interaction strength $c=2/a_{1D}$ as $g_{1D}=-\hbar^{2}c/m$ where
$a_{1D}$ is the effective 1D scattering length. For repulsive
fermions, $c>0$ and for attractive fermions, $c<0$. We shall assume
that the system has periodic boundary conditions i.e.,
$\psi(x_{1},\ldots,x_{i},\ldots,x_{N})=\psi(x_{1},\ldots,x_{i}+L,\ldots,x_{N})$
where $x_{i}$ is the position of the $i$-th particle. There will be
$\kappa$ possible hyperfine states $|1\rangle, |2\rangle, \ldots,
|\kappa\rangle$ that the fermions can occupy. The number of fermions
in the state $|i\rangle$ is given by $N^{i}$ while the Zeeman energy
$\epsilon^{i}_{Z}$ is determined by the magnetic moment
$\mu^{i}_{B}$ and the magnetic field $B$. For brevity, we shall
choose the dimensionless units $\hbar=2m=1$ for the rest of this
paper.

\begin{figure}
\includegraphics[height=130mm]{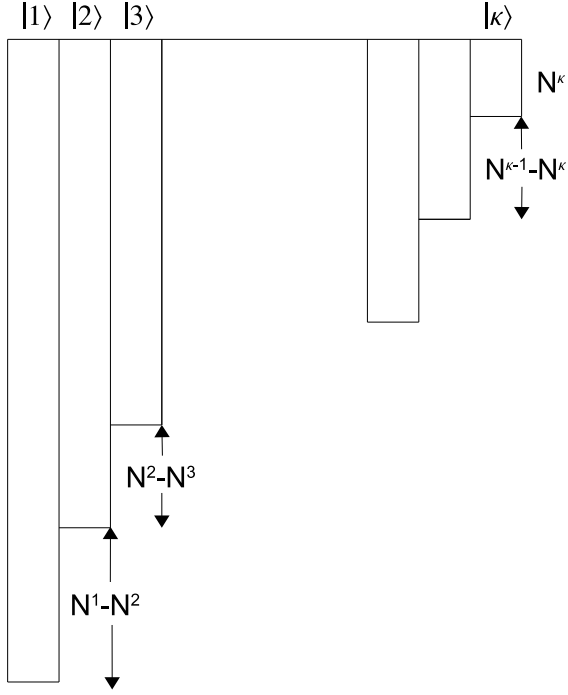}
\vspace{-35mm}\caption{The Young tableau which corresponds to the
irreducible representation
$R_{\psi}=[\kappa^{N^{\kappa}},(\kappa-1)^{N^{\kappa-1}-N^{\kappa}},\ldots,2^{N^{2}-N^{3}},1^{N^{1}-N^{2}}]$.}
\label{fig:young}
\end{figure}

The wavefunction $\psi$ for this Hamiltonian satisfies the symmetry
of an irreducible representation
$R_{\psi}=[\kappa^{N^{\kappa}},(\kappa-1)^{N^{\kappa-1}-N^{\kappa}},\ldots,2^{N^{2}-N^{3}},1^{N^{1}-N^{2}}]$,
where the $N^{i}$s are such that $N^{1}\geq N^{2}\geq\ldots\geq
N^{\kappa}$. The Young tableau which corresponds to this irreducible
representation is given in FIG. \ref{fig:young}. This system has
$SU(\kappa)$ spin symmetry and $U(1)$ charge symmetry. Sutherland
\cite{Sutherland1968} showed that this problem can be solved by
repeatedly using the generalized Bethe's hypothesis which was
introduced by Yang \cite{Yang1967}. To obtain the total momentum
$p=\sum_{j=1}^{N}k_{j}$ and the energy eigenspectrum
$E=\sum_{j=1}^{N}k_{j}^{2}$ for the system, we need to find a set of
quasimomenta $k_{j}$ that satisfies the equations
\begin{equation}
\exp(ik_{j}L)=\prod_{l=1}^{M_{1}}\frac{k_{j}-\Lambda^{(1)}_{l}+ic'}{k_{j}-\Lambda^{(1)}_{l}-ic'}
\qquad j=1,\ldots,N \label{BA1}
\end{equation}
\begin{eqnarray}
\nonumber\prod_{\beta=1}^{M_{p-1}}\frac{\Lambda_{\alpha}^{(p)}-\Lambda_{\beta}^{(p-1)}+ic'}
{\Lambda_{\alpha}^{(p)}-\Lambda_{\beta}^{(p-1)}-ic'}=
-\prod_{\gamma=1}^{M_{p}}\frac{\Lambda_{\alpha}^{(p)}-\Lambda_{\gamma}^{(p)}+2ic'}
{\Lambda_{\alpha}^{(p)}-\Lambda_{\gamma}^{(p)}-2ic'}
\prod_{\delta=1}^{M_{p+1}}\frac{\Lambda_{\alpha}^{(p)}-\Lambda_{\delta}^{(p+1)}-ic'}
{\Lambda_{\alpha}^{(p)}-\Lambda_{\delta}^{(p+1)}+ic'} \\
\alpha=1,\ldots,M_{p} \qquad \textrm{and} \qquad p=1,\ldots,\kappa-1
\label{BA2}
\end{eqnarray}
where $M_{0} := N$, $M_{\kappa}=0$, $\Lambda_{j}^{(0)} := k_{j}$ and 
$\Lambda_{j}^{(\kappa)}$ is undefined. The rapidities
$\Lambda_{j}^{(p)}$ characterize the internal spin degrees of
freedom. The quantum numbers $M_{i}$ are defined as
$M_{i}=\sum_{j=i}^{\kappa-1}N^{j+1}$ and $c'=c/2$. This set of
$\kappa$ coupled algebraic equations are called the BA equations.

\section{The root pattern}
We shall only consider the strong coupling regime where $L|c|\gg 1$.
For the repulsive case, it is easily shown that the roots $k_{j}$
must be real \cite{Takahashi}. However the rapidities
$\Lambda_{j}^{(p)}$ are allowed to take on nonzero imaginary values.
It was first suggested by Lai \cite{Lai1971} that the rapidities
appear as strings in the complex plane of the form
\begin{equation}
\Lambda_{\alpha}^{(p),n,j}=\Lambda_{\alpha}^{(p),n}+i(n+1-2j)|c'|+O(\exp(-\delta
L)) \qquad j=1,\ldots,n \label{lambdaRoots}
\end{equation}
in the thermodynamic limit where $N,L\rightarrow\infty$ while
keeping the ratio $N/L$ fixed. Here $n$ is the length of the string,
$\alpha$ labels each individual string and
$\Lambda_{\alpha}^{(p),n}$ denotes the real part of each string.
Every string must be symmetric along the real axis, i.e. any complex
solution is accompanied by its complex conjugate pair. This solution
for the rapidities hold as long as $L|c|\gg 1$. In other words, it
holds up to order $\exp(-\delta L)$ where $\delta$ is a positive
monotonic increasing function of the interaction strength $|c|$.
When the system is in its ground state, the rapidities do not form
strings. The strings also obey the relation
$M_{p}=\sum_{n=1}^{\infty}nM_{n}^{(p)}$ where $M_{n}^{(p)}$ is the
number of $\Lambda_{j}^{(p)}$ strings with length $n$.

\begin{figure}
\includegraphics[height=100mm]{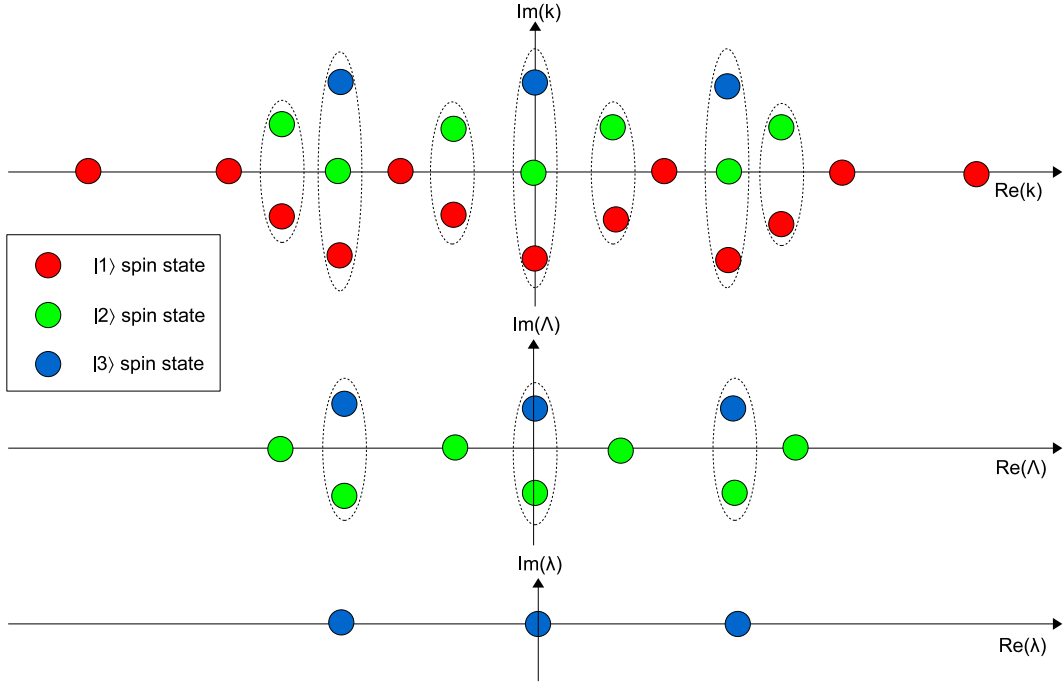}
\caption{(Color online) Root pattern for the $\kappa=3$ case with
$N=23$, $N^{1}=13$, $N^{2}=7$ and $N^{3}=3$ in the ground state
($T=0$). Bounds states are formed for charge and spin rapidities and
are indicated by the dashed boundaries. Accompanying bound states in
each subsequent $\Lambda^{(p)}$-space share the same real part. The
maximum length of any bound state in $k$-space is $\kappa=3$. In an
excited state ($T>0$), strings can also form in
$\Lambda^{(p)}$-space.} \label{fig:roots}
\end{figure}

In the attractive regime, it is found that complex string solutions
of $k_{j}$ also satisfy the BA equations. A system with
$\kappa$-components of fermions has $SU(\kappa)$ symmetry. The
quasimomenta $k_{j}$ may appear as bound states of up to length
$\kappa$. For convenience, we denote the number of $k_{j}$ bound
states with length $1\leq i\leq\kappa$ as $N_{i}$. In the ground
state, none of the bound states can be broken which means that
$N_{i}=N^{i}-N^{i+1}$. A bound state in $k$-space of length $m$ will
take the form
\begin{equation}
k_{\alpha}^{m,j}=\lambda_{\alpha}^{(m-1)}+i(m+1-2j)|c'|+O(\exp(-\delta
L)) \qquad j=1,\ldots,m \label{attractiveroots}
\end{equation}
with real part $\lambda_{\alpha}^{(m-1)}$. In general, a
$k_{\alpha}$ bound state of length $m$ will be accompanied by a
$\Lambda_{\alpha}^{(1)}$ bound state of length $m-1$, a
$\Lambda_{\alpha}^{(2)}$ bound state of length $m-2$ and so on until
a $\Lambda_{\alpha}^{(m-1)}$ bound state of length $1$. Each
accompanying bound state in $\Lambda^{(1)}$-space,
$\Lambda^{(2)}$-space, \ldots, $\Lambda^{(m-1)}$-space will share
the same real part $\lambda_{\alpha}^{(m-1)}$. A graphical depiction
is given in FIG. \ref{fig:roots}.

However, strings in $\Lambda^{(p)}$-space do not have to be
accompanied by any shorter string. Therefore, the difference between
our definition of a bound state and a string is that a bound state
is a string that originates from $k$-space, and is accompanied by
corresponding strings in each subsequent $\Lambda^{(p)}$-space. On
the other hand a string in $\Lambda^{(p)}$-space characterizes the
spin excitations that can exist independently in spin sectors.

\section{The TBA equations: Repulsive case}
The TBA equations which are expressed in the form of dressed
energies allow us to precisely derive numerous thermodynamic
quantities and to analyze the behavior of phase transitions at
critical points. Explicit expressions for the free energy, grand
partition function, pressure, chemical potential and so on can be
directly obtained from the TBA equations. Physically, the TBA
equations give us the energy of elementary excitations above the
ground state. There are several steps to take in order to derive
these equations. We will give an outline of each step in deriving
these equations for the repulsive case, all of which follow from
Yang and Yang's pioneering work \cite{YangYang1969}.

The string solution for $\Lambda_{j}^{(p)}$ gives us a different
form of the BA equations when substituted into the original
equations (\ref{BA1}) and (\ref{BA2}). After lengthy calculations,
we obtain
\begin{equation}
\exp(ik_{j}L)=\prod_{n=1}^{\infty}\prod_{\alpha=1}^{M_{n}^{(1)}}
\frac{k_{j}-\Lambda_{\alpha}^{(1),n}+inc'}{k_{j}-\Lambda_{\alpha}^{(1),n}-inc'}
\end{equation}
\begin{equation}
\prod_{l=1}^{N}\frac{\Lambda_{\alpha}^{(1),n}-k_{l}+inc'}{\Lambda_{\alpha}^{(1),n}-k_{l}-inc'}
\prod_{m=1}^{\infty}\prod_{\beta=1}^{M_{m}^{(2)}}F_{nm}
\left(\Lambda_{\alpha}^{(1),n}-\Lambda_{\beta}^{(2),m}\right)
=-\prod_{m=1}^{\infty}\prod_{\beta=1}^{M_{m}^{(1)}}E_{nm}
\left(\Lambda_{\alpha}^{(1),n}-\Lambda_{\alpha}^{(1),m}\right)
\end{equation}
\begin{eqnarray}
\nonumber
\lefteqn{-\prod_{m=1}^{\infty}\prod_{\beta=1}^{M_{m}^{(p)}}E_{nm}
\left(\Lambda_{\alpha}^{(p),n}-\Lambda_{\beta}^{(p),m}\right)=} \\
&& \prod_{m=1}^{\infty}\prod_{\beta=1}^{M_{m}^{(p-1)}}F_{nm}
\left(\Lambda_{\alpha}^{(p),n}-\Lambda_{\beta}^{(p-1),m}\right)
\prod_{m=1}^{\infty}\prod_{\beta=1}^{M_{m}^{(p+1)}}F_{nm}
\left(\Lambda_{\alpha}^{(p),n}-\Lambda_{\beta}^{(p+1),m}\right)
\end{eqnarray}
\begin{equation}
\prod_{m=1}^{\infty}\prod_{\beta=1}^{M_{m}^{(\kappa-2)}}F_{nm}
\left(\Lambda_{\alpha}^{(\kappa-1),n}-\Lambda_{\beta}^{(\kappa-2),m}\right)
=-\prod_{m=1}^{\infty}\prod_{\beta=1}^{M_{m}^{(\kappa-1)}}E_{nm}
\left(\Lambda_{\alpha}^{(\kappa-1),n}-\Lambda_{\beta}^{(\kappa-1),m}\right).
\end{equation}
The functions $E_{nm}(x)$ and $F_{nm}(x)$ are
\begin{equation}
E_{nm}(x)=\left\{
            \begin{array}{ll}
              e_{|n-m|}(x)e^{2}_{|n-m|+2}(x)\ldots e^{2}_{n+m-2}(x)e_{n+m}(x), & \hbox{for $n\neq m$;}
\\
              e^{2}_{2}(x)e^{2}_{4}(x)\ldots e^{2}_{2n-2}(x)e_{2n}(x), & \hbox{for $n=m$.}
            \end{array}
          \right.
\end{equation}
\begin{equation}
F_{nm}(x)=\left\{
            \begin{array}{ll}
              e_{|n-m|+1}(x)e_{|n-m|+3}(x)\ldots e_{n+m-3}(x)e_{n+m-1}(x), & \hbox{for $n\neq m$;}
\\
              e_{1}(x)e_{3}(x)\ldots e_{2n-3}(x)e_{2n-1}(x), & \hbox{for $n=m$.}
            \end{array}
          \right.
\end{equation}
where
\begin{equation}
e(x)=\frac{x+inc'}{x-inc'}.
\end{equation}

Taking the logarithm of each equation yields
\begin{equation}
k_{j}L+\sum_{n=1}^{\infty}\sum_{\alpha=1}^{M_{n}^{(1)}}\theta\left(\frac{k_{j}-\Lambda_{\alpha}^{(1),n}}{nc'}\right)
=2\pi I_{j} \label{Log1}
\end{equation}
\begin{equation}
\sum_{l=1}^{N}\theta\left(\frac{\Lambda_{\alpha}^{(1),n}-k_{l}}{nc'}\right)
+\sum_{m=1}^{\infty}\sum_{\beta=1}^{M_{m}^{(2)}}\Gamma_{nm}
\left(\frac{\Lambda_{\alpha}^{(1),n}-\Lambda_{\beta}^{(2),m}}{c'}\right)=2\pi
J_{\alpha}^{(1),n}
+\sum_{m=1}^{\infty}\sum_{\beta=1}^{M_{m}^{(1)}}\Theta_{nm}
\left(\frac{\Lambda_{\alpha}^{(1),n}-\Lambda_{\beta}^{(1),m}}{c'}\right)
\label{Log2}
\end{equation}
\begin{eqnarray}
\nonumber\lefteqn{\sum_{m=1}^{\infty}\sum_{\beta=1}^{M_{m}^{(p-1)}}\Gamma_{nm}
\left(\frac{\Lambda_{\alpha}^{(p),n}-\Lambda_{\beta}^{(p-1),m}}{c'}\right)
+\sum_{m=1}^{\infty}\sum_{\beta=1}^{M_{m}^{(p+1)}}\Gamma_{nm}
\left(\frac{\Lambda_{\alpha}^{(p),n}-\Lambda_{\beta}^{(p+1),m}}{c'}\right)}
\\ && =2\pi J_{\alpha}^{(p),n}
+\sum_{m=1}^{\infty}\sum_{\beta=1}^{M_{m}^{(p)}}\Theta_{nm}
\left(\frac{\Lambda_{\alpha}^{(p),n}-\Lambda_{\beta}^{(p),m}}{c'}\right)
\label{Log3}
\end{eqnarray}
\begin{equation}
\sum_{m=1}^{\infty}\sum_{\beta=1}^{M_{m}^{(\kappa-2)}}\Gamma_{nm}
\left(\frac{\Lambda_{\alpha}^{(\kappa-1),n}-\Lambda_{\beta}^{(\kappa-2),m}}{c'}\right)
=2\pi J_{\alpha}^{(\kappa-1),n}+
\sum_{m=1}^{\infty}\sum_{\beta=1}^{M_{m}^{(\kappa-1)}}\Theta_{nm}
\left(\frac{\Lambda_{\alpha}^{(\kappa-1),n}-\Lambda_{\beta}^{(\kappa-1),m}}{c'}\right)
\label{Log4}
\end{equation}
where $I_{j}$ and $J_{\alpha}^{(p),n}$ are odd or half-odd integers
depending on the quantum numbers. The functions $\Theta_{nm}(x)$ and
$\Gamma_{nm}(x)$ are
\begin{equation}
\Theta_{nm}(x)=\left\{
                \begin{array}{ll}
                  \theta\left(\frac{x}{|n-m|}\right)+2\theta\left(\frac{x}{|n-m|+2}\right)
                  +\ldots+2\theta\left(\frac{x}{n+m-2}\right)+\theta\left(\frac{x}{n+m}\right), & \hbox{for $n\neq m$;}
                  \\
                  2\theta\left(\frac{x}{2}\right)+2\theta\left(\frac{x}{4}\right)
                  +\ldots+2\theta\left(\frac{x}{2n-2}\right)+\theta\left(\frac{x}{2n}\right), & \hbox{for $n=m$.}
                \end{array}
              \right.
\end{equation}
\begin{equation}
\Gamma_{nm}(x)=\left\{
                 \begin{array}{ll}
                   \theta\left(\frac{x}{|n-m|+1}\right)+\theta\left(\frac{x}{|n-m|+3}\right)+\ldots+
                   \theta\left(\frac{x}{n+m-3}\right)+\theta\left(\frac{x}{n+m-1}\right), & \hbox{for $n\neq m$;}
\\
                   \theta(x)+\theta\left(\frac{x}{3}\right)+\ldots+\theta\left(\frac{x}{2n-3}\right)
+\theta\left(\frac{x}{2n-1}\right), & \hbox{for $n=m$.}
                 \end{array}
               \right.
\end{equation}
where
\begin{equation}
\theta(x)=2\tan^{-1}(x).
\end{equation}

Let us now introduce continuous monotonic increasing functions
$f(k)=2\pi I_{i}/L$ and $g_{n}^{(p)}(k)=2\pi J_{\alpha}^{(p),n}/L$.
Denote $\rho(k)$ and $\rho^{h}(k)$ as the densities of ``particles''
and ``holes'' in $k$-space. Similarly, denote $\sigma_{n}^{(p)}$ and
$\sigma_{n}^{(p),h}$ as the densities of ``particles'' and ``holes''
for strings with length $n$ in the $p$-th rapidity space. We shall
express equations (\ref{Log1}) to (\ref{Log4}) in the continuum
limit as
\begin{equation}
f(k)=k+\sum_{n=1}^{\infty}\int\theta\left(\frac{k-k'}{nc'}\right)\sigma_{n}^{(1)}(k')dk'
\end{equation}
\begin{eqnarray}
\nonumber g_{n}^{(1)}(k) &=&
\int\theta\left(\frac{k-k'}{nc'}\right)\rho(k')dk'+\sum_{m=1}\int\Gamma_{nm}\left(\frac{k-k'}{c'}\right)\sigma_{m}^{(2)}(k')dk'
\\ &&
-\sum_{m=1}\int\Theta_{nm}\left(\frac{k-k'}{c'}\right)\sigma_{m}^{(1)}(k')dk'
\end{eqnarray}
\begin{eqnarray}
\nonumber g_{n}^{(p)}(k) &=&
\sum_{m=1}^{\infty}\int\Gamma_{nm}\left(\frac{k-k'}{c'}\right)\sigma_{m}^{(p-1)}(k')dk'
+\sum_{m=1}^{\infty}\int\Gamma_{nm}\left(\frac{k-k'}{c'}\right)\sigma_{m}^{(p+1)}(k')dk'
\\ &&
-\sum_{m=1}^{\infty}\int\Theta_{nm}\left(\frac{k-k'}{c'}\right)\sigma_{m}^{(p)}(k')dk'
\end{eqnarray}
\begin{equation}
g_{n}^{(\kappa-1)}(k)=\sum_{m=1}^{\infty}\int\Gamma_{nm}\left(\frac{k-k'}{c'}\right)\sigma_{m}^{(\kappa-2)}(k')dk'
-\sum_{m=1}^{\infty}\int\Theta_{nm}\left(\frac{k-k'}{c'}\right)\sigma_{m}^{(\kappa-1)}(k')dk'.
\end{equation}

Using the relations $\frac{d}{dk}f(k)=2\pi(\rho(k)+\rho^{h}(k))$ and
$\frac{d}{dk}g_{n}^{(p)}(k)=2\pi(\sigma_{n}^{(p)}(k)+\sigma_{n}^{(p),h}(k))$
in the thermodynamic limit, we then have
\begin{equation}
\rho(k)+\rho^{h}(k)=\frac{1}{2\pi}+\sum_{n=1}^{\infty}a_{n}\ast\sigma_{n}^{(1)}(k)
\end{equation}
\begin{equation}
\sigma_{n}^{(1)}(k)+\sigma_{n}^{(1),h}(k)=a_{n}\ast\rho(k)+\sum_{m=1}^{\infty}S_{nm}\ast\sigma_{m}^{(2)}(k)
-\sum_{m=1}^{\infty}T_{nm}\ast\sigma_{m}^{(1)}(k)
\end{equation}
\begin{equation}
\sigma_{n}^{(p)}(k)+\sigma_{n}^{(p),h}(k)=\sum_{m=1}^{\infty}S_{nm}\ast\sigma_{m}^{(p-1)}(k)
+\sum_{m=1}^{\infty}S_{nm}\ast\sigma_{m}^{(p+1)}(k)-\sum_{m=1}^{\infty}T_{nm}\ast\sigma_{m}^{(p)}(k)
\end{equation}
\begin{equation}
\sigma_{n}^{(\kappa-1)}(k)+\sigma_{n}^{(\kappa-1),h}(k)=\sum_{m=1}^{\infty}S_{nm}\ast\sigma_{m}^{(\kappa-2)}(k)
-\sum_{m=1}^{\infty}T_{nm}\ast\sigma_{m}^{(\kappa-1)}(k)
\end{equation}
where the convolution integral $f\ast g(t)=\int
f(\tau)g(t-\tau)d\tau$ is used. The functions $T_{nm}(x)$ and
$S_{nm}(x)$ are
\begin{equation}
T_{nm}(x)=\left\{
            \begin{array}{ll}
              a_{|m-n|}(x)+2a_{|m-n|+2}(x)+\ldots+2a_{m+n-2}(x)+a_{m+n}(x), & \hbox{for $n\neq m$;}
\\
              2a_{2}(x)+2a_{4}(x)+\ldots+2a_{2n-2}(x)+a_{2n}(x), & \hbox{for $n=m$.}
            \end{array}
          \right.
\end{equation}
\begin{equation}
S_{nm}=\left\{
         \begin{array}{ll}
           a_{|m-n|+1}(x)+a_{|m-n|+3}(x)+\ldots+a_{m+n-3}(x)+a_{m+n-1}(x), & \hbox{for $n\neq m$;}
\\
           a_{1}(x)+a_{3}(x)+\ldots+a_{2n-3}(x)+a_{2n-1}(x), & \hbox{for $n=m$.}
         \end{array}
       \right.
\end{equation}
where
\begin{equation}
a_{n}(x)=\frac{1}{\pi}\frac{n|c'|}{(n|c'|)^{2}+x^{2}}.
\end{equation}

The Gibbs free energy per unit length is given by $\Omega/L=E/L-\mu
N/L-TS/L+E_{z}/L$ where the first term corresponds to the ground
state energy, the second term corresponds to the addition and
extraction of particles from the system, the third term represents
the entropy of the system and the last term is associated with the
Zeeman energy per unit length of the system. $\mu$ is the chemical
potential, $T$ is the temperature and $S/L$ is the entropy per unit
length. For each infinitesimal interval $dk$, the energy is
degenerate. The total number of microstates with this energy
degeneracy is given by
\begin{equation}
dW=\frac{(L\rho(k)
dk+L\rho^{h}(k)dk)!}{(L\rho(k)dk)!(L\rho^{h}(k)dk)!}
\times\prod_{p=1}^{\kappa-1}\prod_{n=1}^{\infty}\frac{(L\sigma_{n}^{(p)}(k)dk+L\sigma_{n}^{(p),h}(k)dk)!}
{(L\sigma_{n}^{(p)}(k)dk)!(L\sigma_{n}^{(p),h}(k)dk)!}.
\end{equation}
Using Sterling's approximation $\ln m!=m\ln m-m$ when $m\gg 1$, the
entropy per unit length of the system follows as
\begin{eqnarray}
\nonumber \frac{S}{L} &=&
\int\left\{(\rho+\rho^{h})\ln(\rho+\rho^{h})-\rho\ln\rho
-\rho^{h}\ln\rho^{h}\right\}dk \\ &&
+\sum_{n=1}^{\infty}\sum_{p=1}^{\kappa-1}\int\left\{(\sigma_{n}^{(p)}+\sigma_{n}^{(p),h})
\ln(\sigma_{n}^{(p)}+\sigma_{n}^{(p),h})-\sigma_{n}^{(p)}\ln\sigma_{n}^{(p)}
-\sigma_{n}^{(p),h}\ln\sigma_{n}^{(p),h}\right\}dk
\end{eqnarray}
where the entropy is defined as $S=\int\ln dW$.

The Zeeman energy can be expressed as
\begin{eqnarray}
\nonumber E_{z} &=& -\sum_{p=1}^{\kappa-1}H_{p}N^{p} \\ &=&
-H_{1}N+\sum_{p=1}^{\kappa-1}(H_{p}-H_{p+1})M_{p}.
\end{eqnarray}
The following identities are required for further derivation
\begin{equation}
\frac{N}{L}=\int\rho(k)dk, \qquad \frac{E}{L}=\int k^{2}\rho(k)dk,
\qquad \frac{M_{p}}{L}=\sum_{n=1}^{\infty}\int
n\sigma_{n}^{(p)}(k)dk.
\end{equation}
In the thermodynamic limit, the Zeeman energy can be written as
\begin{equation}
\frac{E_{z}}{L}=-H_{1}\int\rho(k)dk+\sum_{p=1}^{\kappa-1}\sum_{n=1}^{\infty}(H_{p}-H_{p+1})\int
n\sigma_{n}^{(p)}(k)dk.
\end{equation}

Minimizing the Gibbs free energy per unit length when the system is
in equilibrium i.e., $d\Omega=0$ finally gives the TBA equations
\begin{equation}
\varepsilon(k)=k^{2}-\mu-H_{1}-T\sum_{n=1}^{\infty}a_{n}\ast\ln\left(1+e^{-\xi_{n}^{(1)}(k)/T}\right)
\label{TBA-repulsive1}
\end{equation}
\begin{eqnarray}
\nonumber \xi_{n}^{(1)}(k) &=&
n(H_{1}-H_{2})-Ta_{n}\ast\ln\left(1+e^{-\varepsilon(k)/T}\right)
+T\sum_{m=1}^{\infty}T_{nm}\ast\ln\left(1+e^{-\xi_{m}^{(1)}(k)/T}\right)
\\ &&
-T\sum_{m=1}^{\infty}S_{nm}\ast\ln\left(1+e^{-\xi_{m}^{(2)}(k)/T}\right)
\label{TBA-repulsive2}
\end{eqnarray}
\begin{eqnarray}
\nonumber \xi_{n}^{(p)}(k) &=&
n(H_{p}-H_{p+1})+T\sum_{m=1}^{\infty}T_{nm}\ast\ln\left(1+e^{-\xi_{m}^{(p)}(k)/T}\right)
-T\sum_{m=1}^{\infty}S_{nm}\ast\ln\left(1+e^{-\xi_{m}^{(p-1)}(k)/T}\right)
\\ &&
-T\sum_{m=1}^{\infty}S_{nm}\ast\ln\left(1+e^{-\xi_{m}^{(p+1)}(k)/T}\right)
\label{TBA-repulsive3}
\end{eqnarray}
\begin{equation}
\xi_{n}^{(\kappa-1)}(k)=nH_{\kappa-1}+T\sum_{m=1}^{\infty}T_{nm}\ast\ln\left(1+e^{-\xi_{m}^{(\kappa-1)}(k)/T}\right)
-T\sum_{m=1}^{\infty}S_{nm}\ast\ln\left(1+e^{-\xi_{m}^{(\kappa-2)}(k)/T}\right)
\label{TBA-repulsive4}
\end{equation}
where we have defined $\rho^{h}(k)/\rho(k)=\exp(\varepsilon(k)/T)$
and
$\sigma_{n}^{(p),h}(k)/\sigma_{n}^{(p)}=\exp(\xi_{n}^{(p)}(k)/T)$.

The pressure per unit length of the system is
\begin{equation}
p=-\frac{\partial\Omega}{\partial
L}=\frac{T}{2\pi}\int\ln\left(1+e^{-\varepsilon(k)/T}\right)dk.
\label{repulsivepressure}
\end{equation}

\section{The TBA equations: Attractive case}
After substituting the complex solutions for the bound states and
strings into the original BA equations (\ref{BA1}) and (\ref{BA2}),
we obtain
\begin{eqnarray}
\nonumber \exp(im\lambda_{j}^{(m)}L) &=&
(-1)^{m-1}\prod_{p=1}^{m-1}\prod_{q=p}^{\kappa}\prod_{l=1}^{N_{q}}
\frac{\lambda_{j}^{(m)}-\lambda_{l}^{(q)}-i(q+m-2p)|c'|}{\lambda_{j}^{(m)}-\lambda_{l}^{(q)}+i(q+m-2p)|c'|}
\\ && \nonumber \times\prod_{q=m+1}^{\kappa}\prod_{l=1}^{N_{q}}\frac{\lambda_{j}^{(m)}-\lambda_{l}^{(q)}-i(q-m)|c'|}
{\lambda_{j}^{(m)}-\lambda_{l}^{(q)}+i(q-m)|c'|} \\ &&
\times\prod_{n=1}^{\infty}\prod_{\alpha=1}^{M_{n}}\frac{\lambda_{j}^{(m)}-\Lambda_{\alpha}^{(m),n}-in|c'|}
{\lambda_{j}^{(m)}-\Lambda_{\alpha}^{(m),n}+in|c'|} \qquad
\textrm{for} \quad m=1,\ldots,\kappa
\end{eqnarray}
\begin{eqnarray}
\nonumber
\lefteqn{-\prod_{m=1}^{\infty}\prod_{\beta=1}^{M_{m}}E_{nm}(\Lambda_{\alpha}^{(p),n}-\Lambda_{\beta}^{(p),m})=}
\\ && \nonumber
\prod_{l=1}^{N_{p}}\frac{\Lambda_{\alpha}^{(p),n}-\lambda_{l}^{(p)}+in|c'|}{\Lambda_{\alpha}^{(p),n}-\lambda_{l}^{(p)}-in|c'|}
\prod_{m=1}^{\infty}\prod_{\beta=1}^{M_{m}}F_{nm}(\Lambda_{\alpha}^{(p),n}-\Lambda_{\beta}^{(p-1),m})
\\ && \times
\prod_{m=1}^{\infty}\prod_{\beta=1}^{M_{m}}F_{nm}(\Lambda_{\alpha}^{(p),n}-\Lambda_{\beta}^{(p+1),m})
\qquad \textrm{for} \quad p=1,\ldots,\kappa-1.
\end{eqnarray}
Taking the logarithm of each equation yields
\begin{eqnarray}
\nonumber m\lambda_{j}^{(m)}L &=& 2\pi
K_{j}^{(m)}+\sum_{p=1}^{m-1}\sum_{q=p}^{\kappa}\sum_{l=1}^{N_{q}}\theta\left(\frac{\lambda_{j}^{(m)}-\lambda_{l}^{(q)}}
{(q+m-2p)|c'|}\right)+\sum_{q=m+1}^{\kappa}\sum_{l=1}^{N_{q}}
\theta\left(\frac{\lambda_{j}^{(m)}-\lambda_{l}^{(q)}}{(q-m)|c'|}\right)
\\ &&
+\sum_{n=1}^{\infty}\sum_{\alpha=1}^{M_{n}}
\theta\left(\frac{\lambda_{j}^{(m)}-\Lambda_{\alpha}^{(m),n}}{n|c'|}\right)
\label{Log5}
\end{eqnarray}
\begin{eqnarray}
\nonumber
\lefteqn{\sum_{l=1}^{N_{p}}\theta\left(\frac{\Lambda_{\alpha}^{(p),n}-\lambda_{l}^{(p)}}{n|c'|}\right)=}
\\ && \nonumber 2\pi
L_{\alpha}^{(p),n}-\sum_{m=1}^{\infty}\sum_{\beta=1}^{M_{m}}\Gamma_{nm}
\left(\frac{\Lambda_{\alpha}^{(p),n}-\Lambda_{\beta}^{(p-1),n}}{|c'|}\right)
-\sum_{m=1}^{\infty}\sum_{\beta=1}^{M_{m}}\Gamma_{nm}
\left(\frac{\Lambda_{\alpha}^{(p),n}-\Lambda_{\beta}^{(p+1),n}}{|c'|}\right)
\\ && +\sum_{m=1}^{\infty}\sum_{\beta=1}^{M_{m}}
\Theta_{nm}\left(\frac{\Lambda_{\alpha}^{(p),n}-\Lambda_{\beta}^{(p),m}}{|c'|}\right).
\label{Log6}
\end{eqnarray}

We introduce the continuous monotonic increasing functions
$h_{m}(k)=2\pi K_{j}^{(m)}/L$ and $j_{n}^{(p)}(k)=2\pi
L_{\alpha}^{(p),n}/L$. Denote $\rho_{m}(k)$ and $\rho_{m}^{h}(k)$ as
the densities of ``particles'' and ``holes'' for the bound states
with length $m$. Similarly, denote $\sigma_{n}^{(p)}$ and
$\sigma_{n}^{(p),h}$ as the densities of ``particles'' and ``holes''
for strings with length $n$ in the $p$-th rapidity space. In the
continuum limit, equations (\ref{Log5}) and (\ref{Log6}) become
\begin{eqnarray}
\nonumber h_{m}(k) &=&
mk-\sum_{p=1}^{m-1}\sum_{q=p}^{\kappa}\int\theta\left(\frac{k-k'}{(q+m-2p)|c'|}\right)\rho_{q}(k')dk'
\\ &&
-\sum_{q=m+1}^{\kappa}\int\theta\left(\frac{k-k'}{(q-m)|c'|}\right)\rho_{q}(k')dk'
-\sum_{n=1}^{\infty}\int\theta\left(\frac{k-k'}{n|c'|}\right)\sigma_{n}^{(m)}(k')dk'
\end{eqnarray}
\begin{eqnarray}
\nonumber j_{n}^{(p)}(k) &=&
\int\theta\left(\frac{k-k'}{n|c'|}\right)\rho_{p}(k')dk'
+\sum_{m=1}^{\infty}\int\Gamma_{nm}\left(\frac{k-k'}{|c'|}\right)\sigma_{m}^{(p-1)}(k')dk'
\\ && +\sum_{m=1}^{\infty}\int\Gamma_{nm}\left(\frac{k-k'}{|c'|}\right)\sigma_{m}^{(p+1)}(k')dk'
-\sum_{m=1}^{\infty}\int\Theta_{nm}\left(\frac{k-k'}{|c'|}\right)\sigma_{m}^{(p)}(k')dk'.
\end{eqnarray}

From the relations
$\frac{d}{dk}h_{m}(k)=2\pi(\rho_{m}(k)+\rho^{h}_{m}(k))$ and
$\frac{d}{dk}j_{n}^{(p)}(k)=2\pi(\sigma_{n}^{(p)}(k)+\sigma_{n}^{(p),h}(k))$
in the thermodynamic limit, we obtain
\begin{equation}
\rho_{m}(k)+\rho^{h}_{m}(k)=\frac{m}{2\pi}-\sum_{p=1}^{m-1}\sum_{q=p}^{\kappa}a_{q+m-2p}\ast\rho_{q}(k)
-\sum_{q=m+1}^{\kappa}a_{q-m}\ast\rho_{q}(k)
-\sum_{n=1}^{\infty}a_{n}\ast\sigma_{n}^{(m)}(k)
\label{attractivedensity1}
\end{equation}
\begin{equation}
\sigma_{n}^{(p)}(k)+\sigma_{n}^{(p),h}(k)=a_{n}\ast\rho_{p}(k)+\sum_{m=1}^{\infty}S_{nm}\ast\sigma_{m}^{(p-1)}(k)
+\sum_{m=1}^{\infty}S_{nm}\ast\sigma_{m}^{(p+1)}(k)-\sum_{m=1}^{\infty}T_{nm}\ast\sigma_{m}^{(p)}(k).
\label{attractivedensity2}
\end{equation}

The Gibbs free energy for the attractive case has the same
expression as the repulsive case. However, the expressions for each
term in the Gibbs free energy is different from the repulsive case.
The total number of microstates with the same energy degeneracy in
the attractive case is given by
\begin{equation}
dW=\prod_{m=1}^{\kappa}\frac{(L\rho_{m}(k)dk+L\rho_{m}^{h}(k)dk)!}{(L\rho_{m}(k)dk)!(L\rho_{m}^{h}(k)dk)!}
\times\prod_{p=1}^{\kappa-1}\prod_{n=1}^{\infty}\frac{(L\sigma_{n}^{(p)}(k)dk+L\sigma_{n}^{(p),h}(k)dk)!}
{(L\sigma_{n}^{(p)}(k)dk)!(L\sigma_{n}^{(p),h}(k)dk)!}.
\end{equation}
The entropy per unit length of the system is
\begin{eqnarray}
\nonumber \frac{S}{L} &=&
\sum_{m=1}^{\kappa}\int\left\{(\rho_{m}+\rho_{m}^{h})\ln(\rho_{m}+\rho_{m}^{h})-\rho_{m}\ln\rho_{m}
-\rho_{m}^{h}\ln\rho_{m}^{h}\right\}dk \\ &&
+\sum_{n=1}^{\infty}\sum_{i=1}^{\kappa-1}\int\left\{(\sigma_{n}^{(i)}+\sigma_{n}^{(i)h})
\ln(\sigma_{n}^{(i)}+\sigma_{n}^{(i)h})-\sigma_{n}^{(i)}\ln\sigma_{n}^{(i)}
-\sigma_{n}^{(i)h}\ln\sigma_{n}^{(i)h}\right\}dk.
\end{eqnarray}

The ground state energy per unit length was given by Takahashi
\cite{Takahashi1970} as
\begin{equation}
\frac{E}{L}=\sum_{m=1}^{\kappa}\int\left(mk^{2}-\frac{m(m^{2}-1)}{3}|c'|^{2}\right)\rho_{m}(k)dk.
\label{attractive-E}
\end{equation}
It can be easily derived by taking the discrete sum
$E=\sum_{j}k_{j}^{2}$ with the roots given in equation
(\ref{attractiveroots}) and then extending it to the continuum
limit. Therefore the Zeeman energy per unit length is
\begin{eqnarray}
\nonumber \frac{E_{z}}{L} &=& -\sum_{m=1}^{\kappa-1}n_{m}H_{m} \\
&=&
-\sum_{m=1}^{\kappa-1}H_{m}\int\rho_{m}(k)dk+\sum_{m=1}^{\kappa-1}\sum_{n=1}^{\infty}(2H_{m}-H_{m-1}-H_{m+1})
\int n\sigma_{n}^{(m)}(k)dk
\end{eqnarray}
where $H_{\kappa}=0$ because we only need $\kappa-1$ independent
parameters to describe the relative distances between the energy
levels of different fermionic species due to Zeeman splitting. Here
we shall denote $n_{m}=N_{m}/L$ for brevity.

Minimizing the Gibbs free energy with respect to deviations in the
various densities yields a set of coupled integral equations. On
introducing the dressed energy terms
$\exp(\varepsilon_{m}(k)/T)=\rho_{m}^{h}(k)/\rho_{m}(k)$ and
$\exp(\xi_{n}^{(p)}(k)/T)=\sigma_{n}^{(p),h}(k)/\sigma_{n}^{(p)}(k)$,
we arrive at the TBA equations for attractive fermions with
arbitrary spin
\begin{eqnarray}
\nonumber \varepsilon_{m}(k) &=&
mk^{2}-m\mu-H_{m}-\frac{m(m^{2}-1)}{3}|c'|^{2}+T\sum_{p=1}^{m-1}\sum_{q=p}^{\kappa}
a_{q+m-2p}\ast\ln\left(1+e^{-\varepsilon_{q}(k)/T}\right) \\ &&
+T\sum_{q=m+1}^{\kappa}a_{q-m}\ast\ln\left(1+e^{-\varepsilon_{q}(k)/T}\right)
-T\sum_{n=1}^{\infty}a_{n}\ast\ln\left(1+e^{-\xi_{n}^{(m)}(k)/T}\right)
\label{TBA-att1}
\end{eqnarray}
\begin{eqnarray}
\nonumber \xi_{n}^{(p)}(k) &=&
n(2H_{p}-H_{p-1}-H_{p+1})+Ta_{n}\ast\ln\left(1+e^{-\varepsilon_{p}(k)/T}\right)
+T\sum_{m=1}^{\infty}T_{nm}\ast\ln\left(1+e^{-\xi_{m}^{(p)}(k)/T}\right)
\\ &&
-T\sum_{m=1}^{\infty}S_{nm}\ast\ln\left(1+e^{-\xi_{m}^{(p-1)}(k)/T}\right)
-T\sum_{m=1}^{\infty}S_{nm}\ast\ln\left(1+e^{-\xi_{m}^{(p+1)}(k)/T}\right).
\label{TBA-att2}
\end{eqnarray}
Take note that from the definition
$\sigma_{n}^{(\kappa)}=\sigma_{n}^{(\kappa)h}=0$ given earlier,
$\xi_{n}^{(\kappa)}(k)$ is undefined. Here $\varepsilon_{m}(k)$ with
$1\leq m\leq\kappa$ are the dressed energies for the bound states of
length $m$ e.g. $\varepsilon_{1}(k)$ is for unpaired fermions,
$\varepsilon_{2}(k)$ is for pairs, $\varepsilon_{3}(k)$ is for
trions and so on. The Fermi level is at $\varepsilon_{m}(Q_{m})=0$
which implies that the bound states of length $m$ are only occupied
with fermions having quasimomenta $-Q_{m}<k<Q_{m}$. There is an
equivalent description for the band fillings of the strings.

The pressure per unit length for the system is
\begin{equation}
p=-\frac{\partial\Omega}{\partial
L}=\sum_{m=1}^{\kappa}\frac{mT}{2\pi}\int
dk\ln\left(1+e^{-\varepsilon_{m}(k)/T}\right).
\label{attractivepressure}
\end{equation}

\section{The ground state: Strong coupling limit}
\subsection{Repulsive case}
In this section we present the thermodynamic properties of the
system in the ground state where $T\rightarrow 0$ and also in the
antiferromagnetic case where $H_{p}=0$ for every
$p=1,2,\ldots,\kappa-1$. In this regime, there are no string
solutions for each level of rapidity because strings only exist in
excited states. Hence the TBA equations simplify to
\begin{eqnarray}
\varepsilon(k) &=& k^{2}-\mu+a_{1}\ast\xi^{(1)}(k) \label{TBA-rep1} \\
\xi^{(1)}(k) &=&
a_{1}\ast\varepsilon(k)-a_{2}\ast\xi^{(1)}(k)+a_{1}\ast\xi^{(2)}(k) \label{TBA-rep2} \\
\xi^{(p)}(k) &=&
-a_{2}\ast\xi^{(p)}(k)+a_{1}\ast\xi^{(p-1)}(k)+a_{1}\ast\xi^{(p+1)}(k) \label{TBA-rep3} \\
\xi^{(\kappa-1)}(k) &=&
-a_{2}\ast\xi^{(\kappa-1)}(k)+a_{1}\ast\xi^{(\kappa-2)}(k).
\label{TBA-rep4}
\end{eqnarray}

In the regime $c\gg 1$, we can estimate
$a_{1}\ast\varepsilon(k)\approx-2\pi Pa_{1}(k)$ from equation
(\ref{repulsivepressure}). Taking the Fourier transform of equations
(\ref{TBA-rep2}) to (\ref{TBA-rep4}) yields the difference equations
\begin{eqnarray}
\widehat{\xi}^{(1)}(\omega) &=& \widehat{S}(\omega)\left[-2\pi
P+\widehat{\xi}^{(2)}(\omega)\right] \\ \widehat{\xi}^{(p)}(\omega)
&=&
\widehat{S}(\omega)\left[\widehat{\xi}^{(p-1)}(\omega)+\widehat{\xi}^{(p+1)}(\omega)\right]
\\ \widehat{\xi}^{(\kappa-1)}(\omega) &=&
\widehat{S}(\omega)\widehat{\xi}^{(\kappa-2)}(\omega)
\end{eqnarray}
where
\begin{equation}
\widehat{S}(\omega)=\frac{1}{2\cosh|\omega c'|}.
\end{equation}

The general solution to this set of difference equations is
\begin{equation}
\widehat{\xi}^{(p)}(\omega)=-\frac{2\pi P\sinh((\kappa-p)|\omega
c'|)}{\sinh(\kappa|\omega c'|)}.
\end{equation}
The inverse Fourier transform for this function is
\begin{equation}
\xi^{(p)}(k)=\frac{\pi P}{\kappa c'}\left(\frac{\sin(\frac{\pi
p}{\kappa})}{\cos(\frac{\pi p}{\kappa})-\cosh(\frac{\pi k}{\kappa
c'})}\right).
\end{equation}
The convolution integral in equation (\ref{TBA-rep1}) can be
decoupled when $c\gg 1$ where it becomes
\begin{equation}
a_{1}\ast\xi^{(1)}(k)\approx\frac{\pi P}{\kappa
c'}\int\frac{1}{\pi}\frac{nc'}{(nc')^{2}+k^{2}}\frac{\sin(\frac{\pi
p}{\kappa})}{\cos(\frac{\pi p}{\kappa})-\cosh(\frac{\pi k}{\kappa
c'})}dk.
\end{equation}

Using Parseval's theorem,
\begin{equation}
\int_{-\infty}^{\infty}f(t)g(t)dt=\frac{1}{2\pi}\int_{-\infty}^{\infty}\widehat{f}(\omega)\widehat{g}(-\omega)d\omega
\end{equation}
we obtain
\begin{eqnarray}
\nonumber a_{1}\ast\xi^{(1)}(k) &=&
-P\int_{-\infty}^{\infty}e^{-|\omega
c'|}\frac{\sinh((\kappa-1)|\omega c'|)}{\sinh(\kappa|\omega
c'|)}d\omega \\ &=& \nonumber
-2P\int_{0}^{\infty}\frac{e^{(\kappa-2)|\omega
c'|}-e^{-\kappa|\omega c'|}}{e^{\kappa|\omega c'|}-e^{-\kappa|\omega
c'|}}d\omega \\ &=& \frac{P}{\kappa
c'}\left(\mathcal{C}+\Psi\left(\frac{1}{\kappa}\right)\right)
\end{eqnarray}
where $\mathcal{C}=0.577\ldots$ is the Euler-Mascheroni constant and
$\Psi(x)$ is the Psi (Digamma) function. The values of
$\Psi(\frac{1}{\kappa})$ for $\kappa=1,2,3$ are
$\Psi(1)=-\mathcal{C}$, $\Psi(\frac{1}{2})=-\mathcal{C}-2\ln 2$ and
$\Psi(\frac{1}{3})=-\mathcal{C}-\frac{3}{2}\ln
3-\frac{\pi}{2\sqrt{3}}$. Therefore equation (\ref{TBA-rep1})
becomes
\begin{equation}
\varepsilon(k)=k^{2}-\mu+\frac{P}{\kappa
c'}\left(\mathcal{C}+\Psi\left(\frac{1}{\kappa}\right)\right).
\end{equation}

Using the conditions $\varepsilon(\pm Q)=0$, $-2\pi
P=\int_{-Q}^{Q}\varepsilon(k)dk$ and $n=\partial P/\partial\mu$
followed by iteration to keep terms up to order 1/c gives the
thermodynamic expressions
\begin{eqnarray}
Q &\approx& \pi
n\left[1+\frac{2}{\kappa\gamma}\left(\mathcal{C}+\Psi\left(\frac{1}{\kappa}\right)\right)\right]
\\
\mu &\approx&
\pi^{2}n^{2}\left[1+\frac{16}{3\kappa\gamma}\left(\mathcal{C}+\Psi\left(\frac{1}{\kappa}\right)\right)\right]
\\
P &\approx&
\frac{2}{3}\pi^{2}n^{2}\left[1+\frac{6}{\kappa\gamma}\left(\mathcal{C}+\Psi\left(\frac{1}{\kappa}\right)\right)\right]
\\
F &\approx&
\frac{1}{3}\pi^{2}n^{3}\left[1+\frac{4}{\kappa\gamma}\left(\mathcal{C}+\Psi\left(\frac{1}{\kappa}\right)\right)\right]
\end{eqnarray}
where $\gamma=c/n$.

\subsection{Attractive case}
Having derived the expression for the densities of bound states in
equation (\ref{attractivedensity1}), we can derive the ground state
energy from equation (\ref{attractive-E}). As with the repulsive
case, there are no string solutions in the ground state. In the
strong coupling regime $|c|\gg 1$, equation
(\ref{attractivedensity1}) simplifies to
\begin{equation}
\rho_{m}(k)=\frac{m}{2\pi}-\sum_{p=1}^{m-1}\sum_{q=p}^{\kappa}\frac{n_{q}}{\pi}\frac{(q+m-2p)|c'|}
{(q+m-2p)^{2}|c'|^{2}+k^{2}}-\sum_{q=m+1}^{\kappa}\frac{n_{q}}{\pi}\frac{(q-m)|c'|}
{(q-m)^{2}|c'|^{2}+k^{2}}.
\end{equation}

An expression for the Fermi points $Q_{m}$ can be derived by first
evaluating the relation $n_{m}=\int_{-Q_{m}}^{Q_{m}}\rho_{m}(k)dk$
which gives
\begin{eqnarray}
\nonumber n_{m}&=&
\frac{mQ_{m}}{\pi}-\sum_{p=1}^{m-1}\sum_{q=p}^{\kappa}\frac{2n_{q}}{\pi}\tan^{-1}\left(\frac{Q_{m}}{(q+m-2p)|c'|}\right)
-\sum_{q=m+1}^{\kappa}\frac{2n_{q}}{\pi}\tan^{-1}\left(\frac{Q_{m}}{(q-m)|c'|}\right)
\\ &\approx& \frac{mQ_{m}}{\pi}\left(1-\sum_{p=1}^{m-1}\sum_{q=p}^{\kappa}\frac{2n_{q}}{m(q+m-2p)|c'|}
-\sum_{q=m+1}^{\kappa}\frac{2n_{q}}{m(q-m)|c'|}\right)
\end{eqnarray}
and then rearranging to obtain
\begin{equation}
Q_{m}=\frac{\pi
n_{m}}{m}\left(1+\sum_{p=1}^{m-1}\sum_{q=p}^{\kappa}\frac{2n_{q}}{m(q+m-2p)|c'|}
+\sum_{q=m+1}^{\kappa}\frac{2n_{q}}{m(q-m)|c'|}\right)+O\left(\frac{1}{|c'|^{2}}\right).
\end{equation}

The ground state energy per unit length (\ref{attractive-E}) is then
given by
\begin{eqnarray}
\nonumber \frac{E}{L} &=&
\sum_{m=1}^{\kappa}\int_{-Q_{m}}^{Q_{m}}\frac{m^{2}k^{2}}{2\pi}dk
-\sum_{m=1}^{\kappa}\sum_{p=1}^{m-1}\sum_{q=p}^{\kappa}\int_{-Q_{m}}^{Q_{m}}\frac{mn_{q}}{\pi}
\frac{(q+m-2p)|c'|k^{2}}{(q+m-2p)^{2}|c'|^{2}+k^{2}}dk
\\ && \nonumber -\sum_{m=1}^{\kappa}\sum_{q=m+1}^{\kappa}\frac{mn_{q}}{\pi}
\frac{(q-m)|c'|k^{2}}{(q-m)^{2}|c'|^{2}+k^{2}}dk
-\sum_{m=1}^{\kappa}\int_{-Q_{m}}^{Q_{m}}\frac{m(m^{2}-1)}{3}|c'|^{2}\rho_{m}(k)dk
\\ &\approx&
\sum_{m=1}^{\kappa}\frac{m^{2}Q_{m}^{3}}{3\pi}-\sum_{m=1}^{\kappa}\sum_{p=1}^{m-1}\sum_{q=p}^{\kappa}
\frac{2mn_{q}Q_{m}^{3}}{3\pi(q+m-2p)|c'|}
-\sum_{m=1}^{\kappa}\sum_{q=m+1}^{\kappa}\frac{2mn_{q}Q_{m}^{3}}{3\pi(q-m)|c'|}
\\ && \nonumber
-\sum_{m=1}^{\kappa}\frac{m(m^{2}-1)}{3}n_{m}|c'|^{2} \\ &=&
\sum_{m=1}^{\kappa}\frac{\pi^{2}n_{m}^{3}}{3m}+\sum_{m=1}^{\kappa}\sum_{p=1}^{m-1}\sum_{q=p}^{\kappa}
\frac{8\pi^{2}n_{m}^{3}n_{q}}{3m^{2}(q+m-2p)|c|}
+\sum_{m=1}^{\kappa}\sum_{q=m+1}^{\kappa}\frac{8\pi^{2}n_{m}^{3}n_{q}}{3m^{2}(q-m)|c|}
\\ && -\sum_{m=1}^{\kappa}\frac{m(m^{2}-1)}{12}n_{m}|c|^{2}+O\left(\frac{1}{|c|^{2}}\right)
\end{eqnarray}
This expression does not include the Zeeman energy, which is equal
to $-\sum_{m=1}^{\kappa}n_{m}H_{m}$. The actual ground state energy
in the presence of an external magnetic field must include this
term. One can easily derive the ground state energy up to arbitrary
orders in $1/c$ by including higher order contributions to the
Taylor expansions of the functions we applied them to. For better
accuracy, we derived the ground state energy that includes terms up
to order $1/c^{2}$ in the more compact form
\begin{equation}
\frac{E}{L}=\sum_{m=1}^{\kappa}\frac{\pi^{2}n_{m}^{3}}{3m}\left(1+\frac{2A_{m}}{|c|}+\frac{3A_{m}^{2}}{|c|^{2}}\right)
-\sum_{m=1}^{\kappa}n_{m}\epsilon_{m}
\end{equation}
where
\begin{equation}
A_{m}=\sum_{p=1}^{m-1}\sum_{q=p}^{\kappa}\frac{4n_{q}}{m(q+m-2p)}+\sum_{q=m+1}^{\kappa}\frac{4n_{q}}{m(q-m)}
\end{equation}
and $\epsilon_{m}$ is the binding energy for a bound state with
length $m$ i.e.,
\begin{equation}
\epsilon_{m}=\frac{m(m^{2}-1)}{12}|c|^{2}.
\end{equation}

Looking back at the first TBA equation (\ref{TBA-att1}) for the
attractive case, we can denote
$\mu_{m}\equiv\mu+\frac{H_{m}}{m}+\frac{\epsilon_{m}}{m}$ as the
effective chemical potentials for the bound states of length $m$. An
expression for the effective chemical potentials can be derived as
\begin{eqnarray}
\nonumber \mu_{\alpha} &=& \frac{1}{\alpha}\frac{\partial}{\partial
n_{\alpha}}\left(\frac{E}{L}+\sum_{m=1}^{\kappa}n_{m}\epsilon_{m}\right)
\\ &=& \nonumber
\frac{\pi^{2}n_{\alpha}^{2}}{\alpha^{2}}\left(1+\frac{2A_{\alpha}}{|c|}+\frac{3A_{\alpha}^{2}}{|c|^{2}}\right)
+\sum_{m=1}^{\kappa}\frac{2\pi^{2}n_{m}^{3}}{3m\alpha|c|}\frac{\partial
A_{m}}{\partial
n_{\alpha}}+\sum_{m=1}^{\kappa}\frac{2\pi^{2}n_{m}^{3}A_{m}}{m\alpha|c|^{2}}\frac{\partial
A_{m}}{\partial n_{\alpha}} \\ &=&
\frac{\pi^{2}n_{\alpha}^{2}}{\alpha^{2}}\left(1+\frac{2A_{\alpha}}{|c|}+\frac{3A_{\alpha}^{2}}{|c|^{2}}\right)
+\frac{\vec{I}\cdot\vec{B}_{\alpha}}{|c|}+\frac{3\vec{A}\cdot\vec{B}_{\alpha}}{|c|^{2}}.
\label{effectiveMu}
\end{eqnarray}
Here we used the vector notation
$\vec{A}=(A_{1},A_{2},\ldots,A_{\kappa})$,
$\vec{B}_{\alpha}=(B_{\alpha}^{1},B_{\alpha}^{2},\ldots,B_{\alpha}^{\kappa})$
and $\vec{I}=(1,1,\ldots,1)$ where
\begin{eqnarray}
\nonumber B_{\alpha}^{m} &=&
\frac{2\pi^{2}n_{m}^{3}}{3m\alpha}\frac{\partial A_{m}}{\partial
n_{\alpha}} \\ &=&
\frac{8\pi^{2}n_{m}^{3}}{3m^{2}\alpha}\left(\frac{\Theta(\alpha-m-1)}{(\alpha-m)}+\sum_{j=1}^{m-1}
\frac{\Theta(\alpha-j)}{(\alpha+m-2j)}\right)
\end{eqnarray}
for $m,\alpha=1,2,\ldots,\kappa$. The function $\Theta(x)$ is the
Heaviside step function with properties $\Theta(x)=0$ when $x<0$ and
$\Theta(x)=1$ when $x\geq 0$. For the special cases of $SU(3)$ and
$SU(4)$ fermions, see refs.~\cite{Guan2008,Guan2009}.

Zeeman splitting can be characterized by the parameters
$\epsilon_{Z}^{m}$ or $H_{m}$. $\epsilon_{Z}^{m}$ is the Zeeman
energy level for the species of fermions in state $|m\rangle$.
$H_{m}$ on the other hand parameterizes the Zeeman energy level for
bound states with length $m$. Both sets of parameters are related by
the expression
\begin{equation}
\sum_{m=1}^{\kappa}\epsilon_{Z}^{m}n^{m}=-\sum_{m=1}^{\kappa-1}H_{m}(n^{m}-n^{m+1}).
\end{equation}
A consistent solution to this relation for all $n^{m}$ is
$\epsilon_{Z}^{1}=-H_{1}$, $\epsilon_{Z}^{\kappa}=H_{\kappa-1}$ and
$\epsilon_{Z}^{m}=H_{m-1}-H_{m}$ for $2\leq m\leq\kappa-1$. If we
denote the difference between the energy levels of fermions in state
$|m+1\rangle$ and $|m\rangle$ as
$\Delta_{m+1,i}=\epsilon_{Z}^{m+1}-\epsilon_{Z}^{m}$, we obtain the
matrix relation
\begin{equation}
\left(
  \begin{array}{c}
    \Delta_{2,1} \\
    \Delta_{3,2} \\
    \Delta_{4,3} \\
     \vdots \\
     \\
    \Delta_{\kappa-1,\kappa-2} \\
    \Delta_{\kappa,\kappa-1} \\
  \end{array}
\right)=\left(
                \begin{array}{ccccccc}
                  2 & -1 &  &  &  &  &  \\
                  -1 & 2 & -1 &  &  &  &  \\
                   & -1& 2 & -1 &  &  &  \\
                   &  &  & \ddots &  &  &  \\
                   &  &  &  &  &  &  \\
                   &  &  &  & -1 & 2 & -1 \\
                   &  &  &  &  & -1 & 2 \\
                \end{array}
              \right)\left(
                       \begin{array}{c}
                         H_{1} \\
                         H_{2} \\
                         H_{3} \\
                          \vdots \\
                          \\
                         H_{\kappa-2} \\
                         H_{\kappa-1} \\
                       \end{array}
                     \right)
\end{equation}
where the blank entries in the upper and lower triangular sections
of the $(\kappa-1)\times(\kappa-1)$ matrix are defined as zero.

A useful relation between $H_{m}$ and the effective chemical
potentials is \cite{Guan2010}
\begin{equation}
H_{m}=m(\mu_{m}-\mu_{\kappa})+\frac{m\epsilon_{\kappa}}{\kappa}-\epsilon_{m}
\label{externalField}
\end{equation}
from the original definition of the effective chemical potential.
The external fields $H_{m}$ can be tuned experimentally to drive the
system between different phases where bound states of different
lengths exist. In the special case called pure Zeeman splitting
where $\Delta_{m+1,m}=\Delta$ for all $m$, the system has three
distinct magnetic phases. The first phase consists of only
$\kappa$-bound fermions when $H_{1}<H_{1}^{c1}$. The second phase
contains a mixture of $\kappa$-bound fermions and unbound fermions
when $H_{1}^{c1}<H_{1}<H_{1}^{c2}$. And the third phase is made up
of only unbound fermions when $H_{1}>H_{1}^{c2}$.

The critical external fields $H_{1}^{c1}$ and $H_{1}^{c2}$ can be
evaluated from equation (\ref{externalField}) by taking $m=1$ and
using the expressions for the effective chemical potentials from
equation (\ref{effectiveMu}). Doing so gives
\begin{eqnarray}
\nonumber H_{1} &=&
n^{2}\left[\frac{\kappa^{2}-1}{12}|\gamma|^{2}+\pi^{2}(m^{z})^{2}\left(1+\frac{8}{\kappa(\kappa-1)|\gamma|}
-\frac{32m^{z}}{3\kappa(\kappa-1)|\gamma|}\right)\right. \\ &&
-\left.\frac{\pi^{2}}{\kappa^{4}}(1-m^{z})^{2}\left(1-\frac{8}{3\kappa(\kappa-1)|\gamma|}+\frac{32m^{z}}{3\kappa(\kappa-1)|\gamma|}
+\frac{16(1-m^{z})}{3\kappa^{2}|\gamma|}\sum_{j=1}^{\kappa-1}\frac{1}{j}\right)\right]
\end{eqnarray}
where $m^{z}$ is the spin normalized magnetization per particle
density i.e., $m^{z}=\frac{2}{n(\kappa-1)}M^{z}_{\mathrm{true}}$.
This means that while $M^{z}_{\mathrm{true}}$ depicts the true
magnetization of the system, $m^{z}$ normalizes it and only takes on
values from $0$ to $1$. The factor $\frac{\kappa-1}{2}$ corresponds
to the species of fermion that has the highest hyperfine spin
because the unbound phase is made up of these fermions only.

The critical field $H_{1}^{c1}$ corresponds to $m^{z}=0$, while
$H_{1}^{c2}$ corresponds to $m^{z}=1$. Substituting these values for
$m^{z}$ into the equation for $H_{1}$ yields the general results
\begin{eqnarray}
H_{1}^{c1} &=&
n^{2}\left[\frac{\kappa^{2}-1}{12}|\gamma|^{2}-\frac{\pi^{2}}{\kappa^{4}}\left(1-\frac{8}{3\kappa(\kappa-1)|\gamma|}+
\frac{16}{3\kappa^{2}|\gamma|}\sum_{j=1}^{\kappa-1}\frac{1}{j}\right)\right]
\\
H_{1}^{c2} &=&
n^{2}\left[\frac{\kappa^{2}-1}{12}|\gamma|^{2}+\pi^{2}\left(1-\frac{8}{3\kappa(\kappa-1)|\gamma|}\right)\right].
\end{eqnarray}

The system has a linear field-dependent magnetization near the
critical points. For a field slightly above $H_{1}^{c1}$, the
magnetization is given by
\begin{equation}
m_{1}^{z}=\frac{\kappa^{4}}{2\pi^{2}n^{2}}(H_{1}-H_{1}^{c1})\left(1+\frac{8}{\kappa(\kappa-1)|\gamma|}
-\frac{8}{\kappa^{2}|\gamma|}\sum_{j=1}^{\kappa-1}\frac{1}{j}\right).
\end{equation}
On the other hand, for a field that is slightly below $H_{1}^{c2}$,
the magnetization is given by
\begin{equation}
m_{2}^{z}=1-\frac{H_{1}^{c2}-H_{1}}{2\pi^{2}n^{2}}\left(1+\frac{8}{\kappa(\kappa-1)|\gamma|}\right).
\end{equation}

\section{Conclusion}

We have presented a thorough derivation of the TBA equations for a
system of 1D multi-component $\delta$-function interacting fermions
in the presence of external magnetic fields. The key results, in terms 
of which the thermodynamic properties are obtained, are
equations (\ref{TBA-repulsive1})--(\ref{TBA-repulsive4}) for the
repulsive case and equations (\ref{TBA-att1})--(\ref{TBA-att2}) for
the attractive case. The form of our TBA equations differs from
those derived by Schlottmann \cite{Schlottmann1993,Schlottmann1994,S-review},
but are nevertheless possibly interchangeable. To see how this can
be done for the $SU(2)$ case, the reader is referred to Takahashi's
book \cite{Takahashi}. 
The nature of charge bound states describing different sizes of atomic 
molecules was studied in
terms of BA root patterns in the attractive regime. Quantum phase
diagrams and quantum phase transitions were analytically studied
from the dressed energy formalism. We also presented the ground
state energies for the strongly repulsive and strongly attractive
regimes. 
We found that all phase transitions for 1D $\delta$-function
attractive fermions are of second order with a linear
field-dependent magnetization in the vicinities of critical points, 
rather than a square-root field-dependent magnetization
\cite{Schlottmann1993,Schlottmann1994,S-review}. 
The linear field-dependence is as found for the SU(2) case \cite{Guan2007} 
and for the Hubbard model \cite{Woyn,WP}.

For the $SU(2)$ case, the TBA equations provide a
comprehensive understanding of FFLO pairing and finite temperature
thermodynamics of Tomonaga-Luttinger liquids \cite{Guan2007,Zhao}. The key
features of phase diagrams and low temperature density profiles of
trapped 1D spin-1/2 fermions were experimentally confirmed by
matching theoretical predictions from the TBA equations, see Liao
{\em et al.} \cite{Liao2009} and references therein. 
For the $SU(\kappa)$ case, these TBA
equations were used to study the universal thermodynamics through
the derivation of the equations of state \cite{Guan2010}. The
results presented in this paper provide the setting for further study of
quantum critical behavior in 1D interacting Fermi gases, where the
exact BA solutions provide insight into the physical origins of
quantum criticality.

\begin{acknowledgments}
This work is supported by the Australian Research Council. MTB and
XWG thank the Institute of Physics, Chinese Academy of Science,
Beijing, China for kind hospitality.
\end{acknowledgments}

\end{document}